# TiO$_2$ nanotubes with different spacing, Fe$_2$O$_3$ decoration and their evaluation for Li-ion battery application

Selda Ozkan[1†], Gihoon Cha[1†], Anca Mazare[1] and Patrik Schmuki[1,2]

[1] *Department of Materials Science and Engineering, WW4-LKO, University of Erlangen-Nuremberg, Martensstrasse 7, D-91058 Erlangen, Germany*

[2] *Chemistry Department, Faculty of Sciences, King Abdulaziz University, 80203 Jeddah, Saudi Arabia Kingdom*

**\*Corresponding author:**

Prof. Dr. Patrik Schmuki:

Tel.: +49-9131-852-7575, Fax: +49-9131-852-7582, Email:schmuki@ww.uni-erlangen.de

†These authors equally contributed to this work.

In present work, we report on the use of organized TiO$_2$ nanotube layers with a regular intertube spacing for the growth of highly defined α-Fe$_2$O$_3$ nano-needles in the interspace. These α-Fe$_2$O$_3$ decorated TiO$_2$ NTs are then explored for Li-ion battery applications and compared to classic close-packed NTs that are both decorated with various amounts of nanoscale α-Fe$_2$O$_3$. We show that nanotubes with tube-to-tube spacing allow a uniform decoration of individual nanotubes with regular arrangements of hematite nano-needles. The tube spacing also facilitates the electrolyte penetration as well as yields better ion diffusion. While bare close-packed NTs show higher capacitance, e.g., 71 µAh cm$^{-2}$ than bare spaced NTs with e.g., 54 µAh cm$^{-2}$, the hierarchical decoration with secondary metal oxide, α-Fe$_2$O$_3$, remarkably enhances the Li-ion battery performance. Namely, spaced nanotubes with α-Fe$_2$O$_3$ decoration have an areal capacitance of 477 µAh cm$^{-2}$, i.e., show up to nearly ~8 times higher capacitance. However, the areal capacitance of close-packed NTs with α-Fe$_2$O$_3$ decoration saturates at 208 µAh cm$^{-2}$, i.e., is limited to ~3 times increase.

# 1. Introduction

Over the last decades, the formation, properties and functionalization of electrochemically formed TiO$_2$ nanotubes (NTs) attracted considerable interest and such structures have been explored for a broad range of applications, e.g. dye-sensitized solar cells [1,2], photocatalysis [2], biomedicine [3,4] and other energy storage devices [5–7]. Meanwhile, a high level of control over the morphological features of the nanotubes, such as diameter, length, wall thickness, wall morphology, as well as over crystal phases (amorphous, anatase, rutile) has been achieved [5,8]. A range of morphological variations has been produced, including bamboo tubes, branched tubes, the fabrication of membranes, or single-walled TiO$_2$ NTs [5,9–12].

Generally, self-ordered TiO$_2$ nanotube arrays grow in a hexagonally close-packed (CP) configuration, i.e., have no or only a minor spacing at the top (top spacing). However, previous research shows that with certain electrolyte compositions, spaced NTs, i.e., showing regular as well as tunable gaps between individual tubes, are obtained [13–16]. This particularly if the organic electrolyte is diethylene glycol (DEG) or dimethyl sulfoxide (DMSO). Few more electrolytes, e.g., ethylene glycol (EG), tri(tetra, poly)-ethylene glycol, were reported to result in the growth of spaced NTs under specific anodization conditions [17,18]. Nanotubes grown on Ti metal, which serves as a back contact, can outperform classic geometries if a conformal/hierarchical decoration is allowed/accommodated [19], as suggested by Simon and Gogotsi for carbon nanotubes [20]. Here, we introduce the use of TiO$_2$ spaced (SP) nanotubular arrays as a scaffold that offers a wide intertube space to be filled/loaded with a secondary active metal oxide for energy storage application, in the form of Li-ion batteries (LIBs).

Among different electrochemical energy conversion and storage technologies, LIBs provide a high energy density, long cycle life, low self-discharge and are environmentally friendly [5,21,22]. LIBs consist of two Li-intercalation electrodes, a negative electrode (anode, e.g., graphite) and a positive electrode (cathode, e.g., LiCoO$_2$). Graphite is the most common anode material, however, as an anode material it has limitations due to a low specific capacity and poor safety (e.g., dendrite formation) [22–24]. TiO$_2$ represents an alternative anode material to graphite, since it possesses the combination of sufficient intercalation capacity with a good cyclic stability (low volume expansion, i.e., <4%), low production costs, and low toxicity [25,26].

Various nano-geometries as anode materials for LIB such as nanowires, nanorods, nanoparticles, mesocages, and nanopores/nanotubes have been used [22,27–30]. Nanotubes/nanopores morphologies (by anodization) have an advantage over the other morphologies as they can be directly grown from the metal in a vertically aligned form that provides electron transport directionality [5,11]. An early attempt of using $TiO_2$ nanotubes in Li-ion batteries was reported by Zhou et al. in 2003 [25], using hydrothermally grown anatase nanotubes with ~300 nm individual tube length and ~8 nm diameter (specific capacitance, 182 mAh/g at 80 mA/g).

The capacitance of $TiO_2$ is however limited due to the low electronic conductivity or the high intrinsic resistance [31]. The most widely used techniques to improve the capacitance, are to modify the nanotubes by hierarchical or conformal decoration with highly pseudocapacitive metal oxides, e.g., $SnO_2$, $Co_3O_4$, $CoO$, $NiO_2$, $Fe_2O_3$, $Nb_2O_5$ [5,32–34]. Amongst the various transition metal oxides, $Fe_2O_3$ is promising owing to its high theoretical capacity (i.e., 1000 mAh $g^{-1}$), high abundance, and low processing costs [33,35]. Particularly interesting is that the nano-size α-$Fe_2O_3$ has better electrochemical performance than the micro-size α-$Fe_2O_3$, as lithium can be inserted in the nanostructure without a phase transformation [33].

In the present work, we grow spaced $TiO_2$ NTs by electrochemical anodization of a Ti foil in di-ethylene glycol (DEG) electrolyte and then hierarchically coat the tubes with α-$Fe_2O_3$ nanogeometries. Such $TiO_2$ nanotubular arrays show a strong improvement of the Li-ion battery performance in comparison to the classic close-packed tube configurations.

## 2. Experimental section

Prior to anodization, 0.1 mm thick Ti foils (99.6% pure tempered annealed, ADVENT) were degreased by sonication in acetone, ethanol, and distilled water, respectively, and dried under a nitrogen ($N_2$) stream. To fabricate spaced nanotubes (NTs), anodization was performed in diethylene glycol (DEG) electrolyte with additions of 2 wt% HF (40%), 7 wt% $H_2O$, and 0.3 wt% $NH_4F$ at 30 V (room temperature) for 10 h. The classic close-packed NTs were formed in 1 M $H_2O$ and 0.1 M $NH_4F$ containing ethylene glycol (EG) electrolyte at an applied voltage of 40 V for 15 min at 50 °C. Anodization was carried out in a two-electrode configuration with Pt as cathode and Ti substrate as anode (IMP-Series Jaissle Potentiostat). After anodization, the samples were immersed in ethanol overnight and dried in a $N_2$ stream. Before the $Fe_2O_3$ decoration, the nanotubular layers were annealed at 450 °C (air) for 2 h using a rapid thermal annealer. Iron (III) chloride hexahydrate ($FeCl_3.6H_2O$) was used as a precursor to prepare 10-160 mM suspensions in de-ionized water. β-FeOOH decoration of the nanotubes was carried out by a chemical precipitation technique at 80 °C in an oven (Heraeus, Germany) for 4 h, according to Eq. 1. Subsequently, nanotubes with β-FeOOH decoration were annealed at 400 °C for 2 h in air to form α-$Fe_2O_3$ [36] according to Eq. 2.

$Fe^{3+} + 3H_2O \rightarrow FeOOH + 3H^+$ 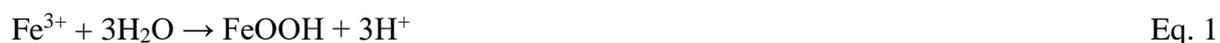 Eq. 1

$$2\text{FeOOH} \rightarrow \text{Fe}_2\text{O}_3 + \text{H}_2\text{O} \qquad \text{Eq. 2}$$

*2.1 Characterization and measurements*

The morphology was characterized using a field-emission SEM (S4800 Hitachi) coupled with an EDX (Genesis 4000). XRD (X'pert PhilipsMPD with a Panalytical X'celerator detector, Germany) was performed using graphite monochromized CuKα radiation (Wavelength 1.54056 Å). The chemical composition was characterized by XPS (PHI 5600, spectrometer, USA) using Al Kα monochromatized radiation (spectra were shifted to C1s a 284.8 eV) and the peaks were fitted with Multipak software.

Electrochemical measurements were conducted in a propylene carbonate electrolyte consisting of 1 M LiClO$_4$ solution at room temperature using a two-electrode setup in Argon filled glove box (MBraun, water and oxygen content below 0.5 ppm). Cyclic voltammetry tests were performed in an electrochemical cell, exposing ≈0.567 cm$^2$ of the sample surface using a two-electrode set-up, i.e. Li metal served as a counter/cathode and the nanotube layer as working electrode/anode. The cyclic voltammetry tests were done in the voltage range of 0–3 V at a scan rate of 1 mV/s (the 5$^{th}$ cycle was considered if not stated otherwise). The galvanostatic charge-discharge tests were conducted using a Swagelok cell that consists of a separator with a working area of ≈0.5 cm$^2$ (i.e., GF/F, Whatman), a Li cathode (Alfa Aesar, Li foil-99.9%), and the tubular layers as an anode.

## 3. Results and discussion

Figure 1 shows SEM images of spaced and close-packed NTs. The spaced NTs are grown in DEG-based electrolyte at 30 V for 10 h (Figure 1 (a1)–(a2)) under optimized conditions (voltage, temperature, electrolyte composition, e.g., HF, H$_2$O and NH$_4$F) and have a slow growth rate. In general, spaced nanotubes grow under a metastable situation and in the voltage range between 10 to 40 V, while higher or lower voltages than this range lead to porous oxide or only sponge oxide structures, as previously reported [37].

Classic close-packed NTs are used as a reference (CP-TiO$_2$) and fabricated in EG-based electrolyte, see Figure 1 (b1)-(b2). The geometrical features of the as-formed nanotubes are as follows: EG close-packed NTs have a tube diameter of ~81 nm, while DEG spaced NTs have a diameter of ~127 nm, and both spaced and close-packed NTs are ~3 µm long. From the top view, spaced NTs have ring-like features and close-packed NTs have grass-like structures at the top due to over-etching of tube tops in the electrolyte. The significant morphological difference between close-packed and spaced NTs is that spaced NTs have a spacing (tube-to-tube) of 150±40 nm between the individual NTs.

Another remarkable difference between close-packed and spaced NTs is their wall morphology i.e., double-walled structure (consists of carbon rich inner shell) or single-walled, see Figure 1 (a2)-(b2). In the case of close-packed NTs, a double-walled structure of the nanotubes is revealed from top to bottom. On the other hand, in the case of spaced tubes, the double-walled structure is visible only close to the bottom. Overall, the spacing in between individual nanotubes and the thin-wall structure of spaced NTs will lead to a larger area available for the decoration with secondary oxide materials.

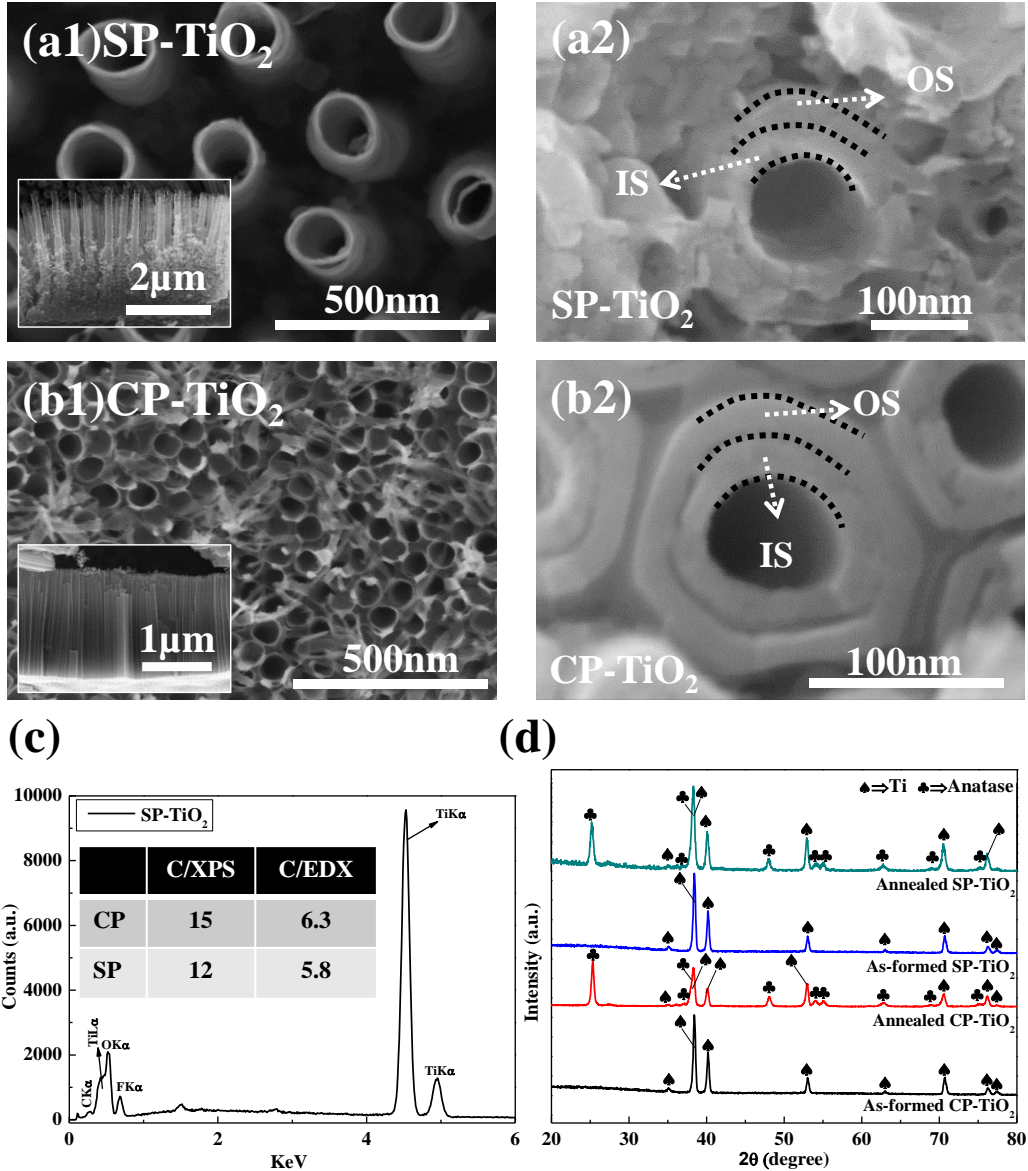

**Figure 1 (a1)** Top-view and inset cross-section SEM image of spaced $TiO_2$ nanotubes (SP-$TiO_2$). **(a2)** Double-wall SP-$TiO_2$ nanotubes, nanotubes cracked close to the bottom. **(b1)** Top-view and inset cross-section SEM image of close-packed $TiO_2$ nanotubes (CP-$TiO_2$). **(b2)**

Double-wall CP-TiO$_2$ nanotubes, nanotubes cracked close to the middle. (**c**) Energy dispersive X-ray spectroscopy (EDX) analysis result for as-formed SP-TiO$_2$ (inset table shows carbon content of as-formed CP-TiO$_2$ and SP-TiO$_2$ measured by XPS and EDX analysis). (**d**) X-ray diffraction (XRD) patterns of as-formed, annealed CP-TiO$_2$ and SP-TiO$_2$ NTs.

Figure 1(c) shows EDX and XPS analysis of spaced and close-packed NTs: both EG and DEG NTs have a higher carbon content at the top most layer. The higher carbon content at the top (in XPS) can be ascribed to the carbon pick-up or adsorption of carbon from the electrolyte. In the case of EDX analysis, CP NTs have a slightly higher carbon content due to carbon rich inner shell from top to bottom.

X-ray diffraction patterns of as-formed and annealed NTs are given in Figure 1(d). After annealing, the anatase-rutile phase composition of structures shows differences as close-packed NTs have an anatase fraction of 92 % (8 % rutile) and spaced NTs have 71 % (29 % rutile). The higher rutile composition of spaced NTs is attributed to the easy conversion of the sponge oxide to rutile. The spongy oxide is concentrated in the lower part of the tubular layer and its amount can increase or decrease depending on the anodization conditions (applied temperature, applied voltage, and water content) [37,38].

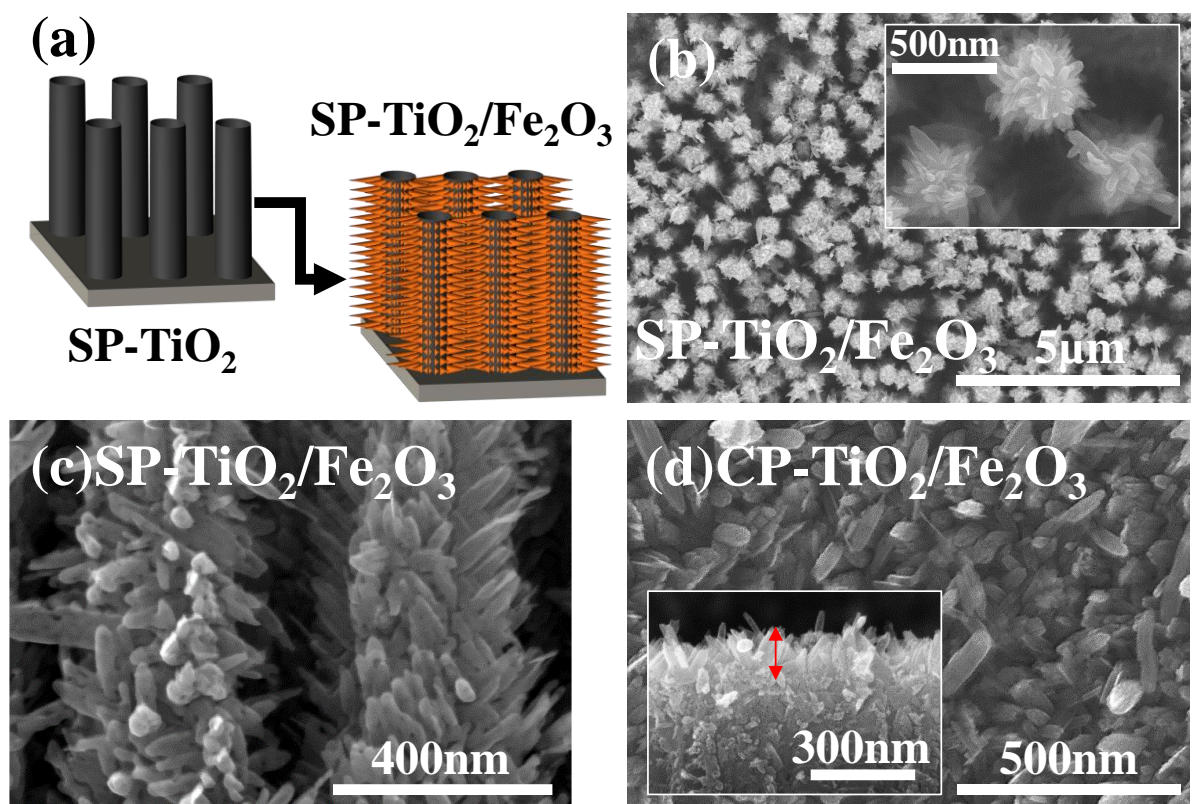

**Figure 2 (a)** Schematic drawing of bare spaced (SP-TiO$_2$) and Fe$_2$O$_3$ decorated spaced TiO$_2$ NTs (SP-TiO$_2$/Fe$_2$O$_3$). (**b**) Top view (inset shows high magnification image) and (**c**) cross-section SEM image of Fe$_2$O$_3$ decorated spaced TiO$_2$ nanotubes (SP-TiO$_2$/Fe$_2$O$_3$). **(d)** Top-view and inset cross-section SEM image of Fe$_2$O$_3$ decorated close-packed TiO$_2$ nanotubes (CP-TiO$_2$/Fe$_2$O$_3$).

The assembly of the hierarchical TiO$_2$/Fe$_2$O$_3$ electrode is performed in two-steps: *i)* self-organizing anodization of Ti to form spaced TiO$_2$ nanotubes (SP-TiO$_2$), and *ii)* decoration of NTs with α-Fe$_2$O$_3$ (SP-TiO$_2$/Fe$_2$O$_3$), as illustrated in Figure 2 and S1-2. α-Fe$_2$O$_3$ decoration is established by using FeCl$_3$.6H$_2$O (in DIW) at 80 °C. The decorated iron is in the form of iron oxyhydroxide and contains Cl$^-$ (see EDX in Figure S3(a)). Afterward the electrode is dried and sintered at 400 °C in air for 2 h to crystallize the iron oxyhydroxide particles according to Eq. 1 and Eq. 2. It is worth mentioning that a similar electrode assembly procedure is followed for close-packed NTs (CP-TiO$_2$/Fe$_2$O$_3$). The obtained α-Fe$_2$O$_3$ has a nano-needle structure with sharp tips, see Figure 2(b)–(d), S1-S2. The nano-needles grow on the walls of NTs from top to the bottom and exist regularly between as well as on the inner walls of the NTs, see Figure S4.

To evaluate the optimum α-Fe$_2$O$_3$ decoration, both close-packed and spaced nanotubes were decorated in solutions with different Fe-precursor concentrations (i.e., 10 to 160 mM). As expected, spaced NTs can accommodate more α-Fe$_2$O$_3$ nano-needles compared with close-packed NTs, i.e., even when the concentration is increased to 160 mM there is no clogging of the tube tops and the α-Fe$_2$O$_3$ nano-needles uniformly cover the outer as well as the inner walls of the NTs from tube top to bottom (see Figure 2(b)–(c) for 120 mM, Figure S1 for different loadings from 40 to 160 mM and Figure S5-6 for elemental mapping). Contrary to spaced NTs, in the case of close-packed NTs, low concentrations, for instance 10 mM, lead to open tube tops, however, at high concentrations, such as ≥40 mM, α-Fe$_2$O$_3$ nano- needles clog the tube openings (see Figure 2(d) and S2).

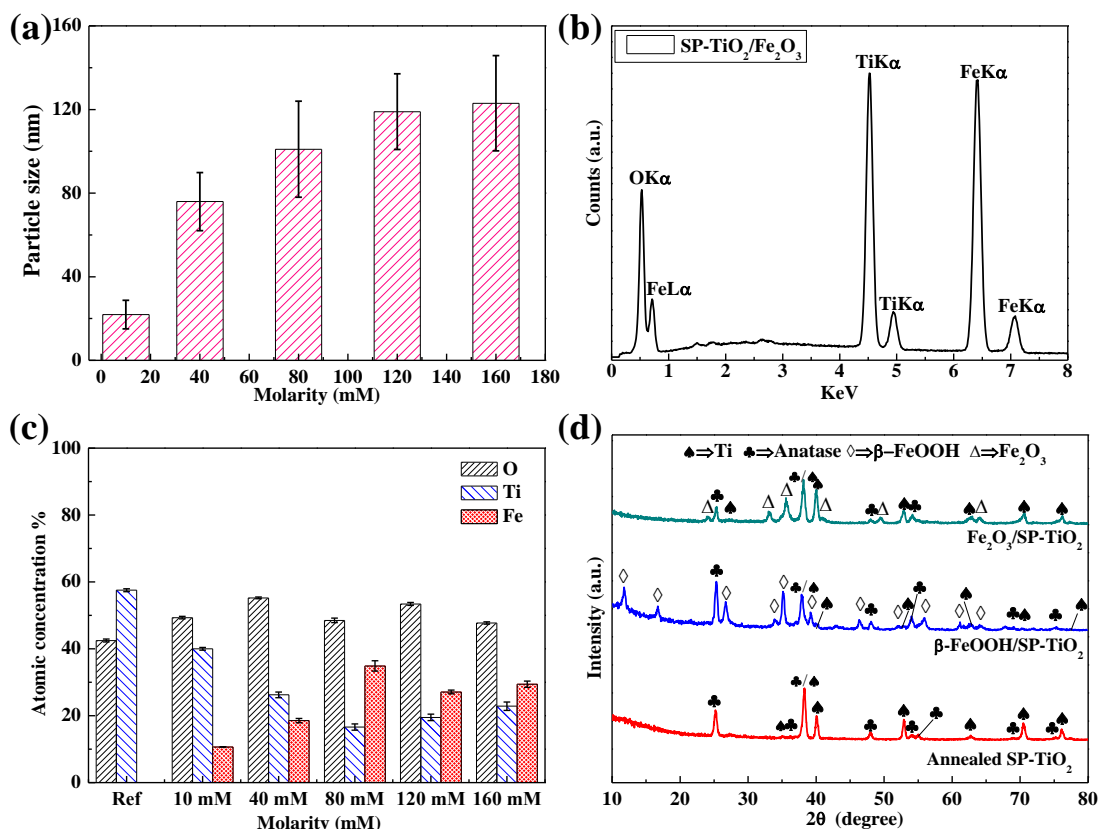

**Figure 3 Spaced nanotubes: (a)** Variation of the $Fe_2O_3$ particle size vs. concentration of the iron precursor used for deposition. **(b)** EDX analysis of $\alpha$-$Fe_2O_3$ decorated spaced $TiO_2$ NTs (SP-$TiO_2$/$Fe_2O_3$). **(c)** Variation of atomic concentration of Ti, Fe, O as a function of the concentration of iron precursor for bare (SP-$TiO_2$) and $\alpha$-$Fe_2O_3$ decorated spaced NTs (SP-$TiO_2$/$Fe_2O_3$) measured by EDX. **(d)** XRD patterns of as-formed, as-deposited $\beta$-FeOOH on SP-$TiO_2$ and $\alpha$-$Fe_2O_3$ on SP-$TiO_2$/ NTs.

The concentration of the iron salt/precursor affects the size of $\alpha$-$Fe_2O_3$ nano-needles, i.e., particle size increases with Fe-precursor concentration as shown in Figure 3(a). Literature reports that particle size of iron (III) oxide is controlled by pH, temperature, the nature of the salts, as well as the concentration and furthermore nano-sized iron oxide particles show better performance than micro-sized [33,39,40].

Additionally, the EDX elemental analyses for close-packed and spaced NTs with/without $\alpha$-$Fe_2O_3$ decoration obtained using various Fe-precursor concentrations are shown in Figure 3(b)-(c) and S3(b). In the case of close-packed NTs, the atomic ratio of Fe reaches 18.1 for 40–80 mM and then it drastically decreases (Figure S3(b)). On the other hand, for spaced NTs, at 80 mM Fe concentration, the highest Fe content (34.2 at. %) is reached and then it slightly decreases, see Figure 3(c).

Figure 3(d) demonstrates X-ray diffraction patterns of spaced nanotubes with/without iron-oxide decoration (see Figure S3(c) for XRD patterns of close-packed NTs with/without α-$Fe_2O_3$ decoration). Annealed $TiO_2$ NTs crystallize in the form of anatase (JCPDS 21-1272) and rutile (JCPDS 21-1276). The as-decorated nano-needles are in the form of β-FeOOH (Akaganeite; Tetragonal crystal form) with peaks located at 11.84° (110), 16.79° (200), 26.72° (310), 35° (211), and 55.9° (521). After annealing, the β-FeOOH is converted to hematite, hexagonal crystal form (α-$Fe_2O_3$, JCPDS No. 89-0596), with peaks located at 24.13° (012), 33.12° (104), 35.62° (110), 49.4° (024) and 63.97° (300).

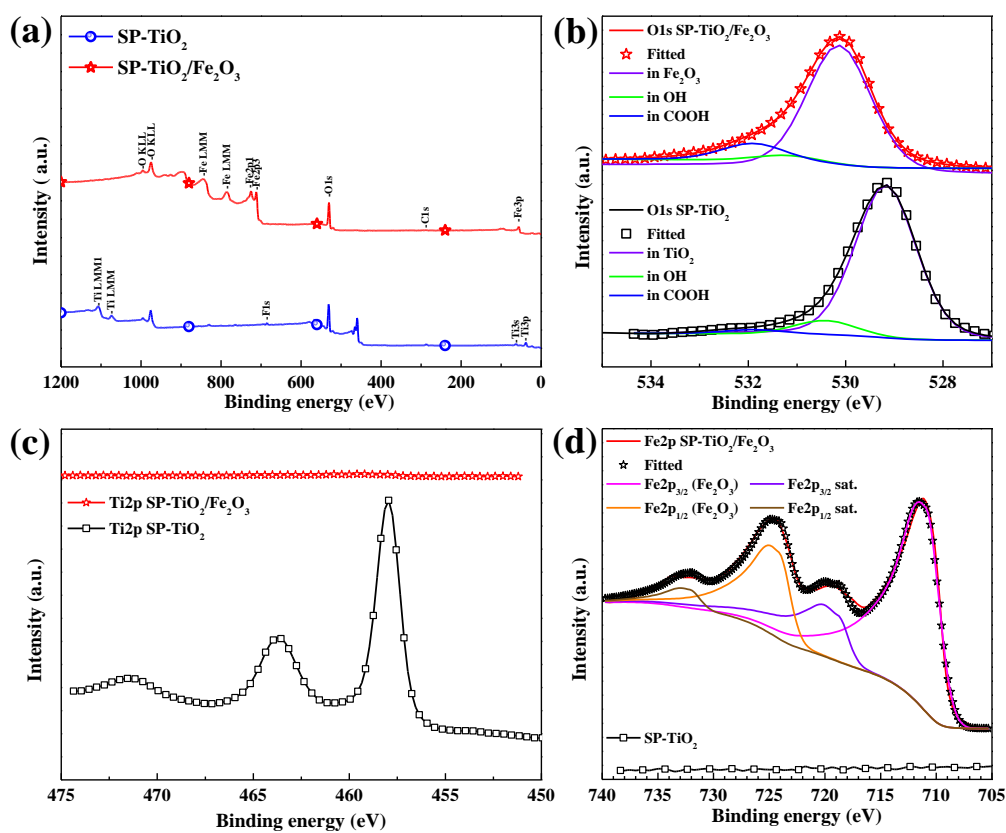

**Figure 4 X-ray photoelectron spectroscopy (XPS) analysis: (a)** XPS survey spectra of bare SP-$TiO_2$ and α-$Fe_2O_3$ decorated spaced (SP-$TiO_2$/$Fe_2O_3$) NTs. (**b**) XPS O1s peaks and fittings for SP-$TiO_2$ and SP-$TiO_2$/$Fe_2O_3$ nanotubes. (**c**) XPS Ti2p peaks for bare SP-$TiO_2$ and SP-$TiO_2$/$Fe_2O_3$ NTs. (**d**) XPS Fe2p peak and fitting for SP-$TiO_2$/$Fe_2O_3$ NTs.

The chemical composition of SP-$TiO_2$ and SP-$TiO_2$/$Fe_2O_3$ tubular layers was evaluated by XPS as shown in Figure 4. The survey spectra of the as-formed nanotubes confirm the presence of Ti, C, O, F, and after the α-$Fe_2O_3$ decoration only Fe, O and C peaks are visible. The high-resolution spectra of O1s (with peak fitting), Ti2p and Fe2p (with peak fitting) are given in Figure 4(b)–(d). The reference sample, that is the bare SP-$TiO_2$ nanotubes, shows the typical

O1s peak consisting of an O signal from $TiO_2$ at ≈529.2 eV, O from OH bonds at ≈530.4eV and at ≈531.9 eV from C-O, as attributed in Figure 4(b), with the corresponding $Ti2p_{3/2}$ peak at ≈458.0 eV. For the α-$Fe_2O_3$ decorated samples, no Ti2p signals are observed at the surface, indicating the uniform coverage of $TiO_2$ tube walls with α-$Fe_2O_3$. The O1s peak can be associated with the O signal from α-$Fe_2O_3$ at ≈530.2 eV (in line with literature [41]), O from OH bonds at ≈531.2 eV and at ≈531.9 eV from C-O. As expected, Fe peaks are detected only for the α-$Fe_2O_3$ decorated samples. The peaks at 710.2 eV ($Fe2p_{3/2}$) and 724.6 eV ($Fe2p_{1/2}$) indicate the presence of α-$Fe_2O_3$ and are consistent with the values reported in literature for the α-$Fe_2O_3$ phase [41–43] – this holds for the binding energy values of $Fe2p_{3/2}$ and $Fe2p_{1/2}$ as well as the clearly distinguishable satellite peaks at ≈719.5eV and at ≈732.4eV.

Such hierarchical structures are further evaluated for Li-ion battery (LIB) application and a comparison of the battery performance of the spaced and the close-packed nanotubes is demonstrated in Figure 5. Cyclic voltammetry (CV) and galvanostatic charge/discharge tests were conducted in a propylene carbonate electrolyte consisting of 1 M $LiClO_4$ solution as described in the experimental section. The Li-ion intercalation into $TiO_2$ lattice occurs according to:

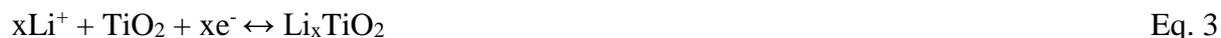

$xLi^+ + TiO_2 + xe^- \leftrightarrow Li_xTiO_2$  Eq. 3

In the CV curves, the $Li^+$ intercalation peak is observed at 1.64 and 1.61 V for SP-$TiO_2$ and CP-$TiO_2$ NTs, and the $Li^+$ deintercalation peak is located at 2.23 V for SP-$TiO_2$ and at 2.31 V for CP-$TiO_2$, respectively. From Figure 5 it is evident that the CV curves of the bare CP-$TiO_2$ NTs reach a higher current density than the SP-$TiO_2$ NTs. This is ascribed to the higher surface area (i.e., higher number of nanotubes) for close-packed tubes.

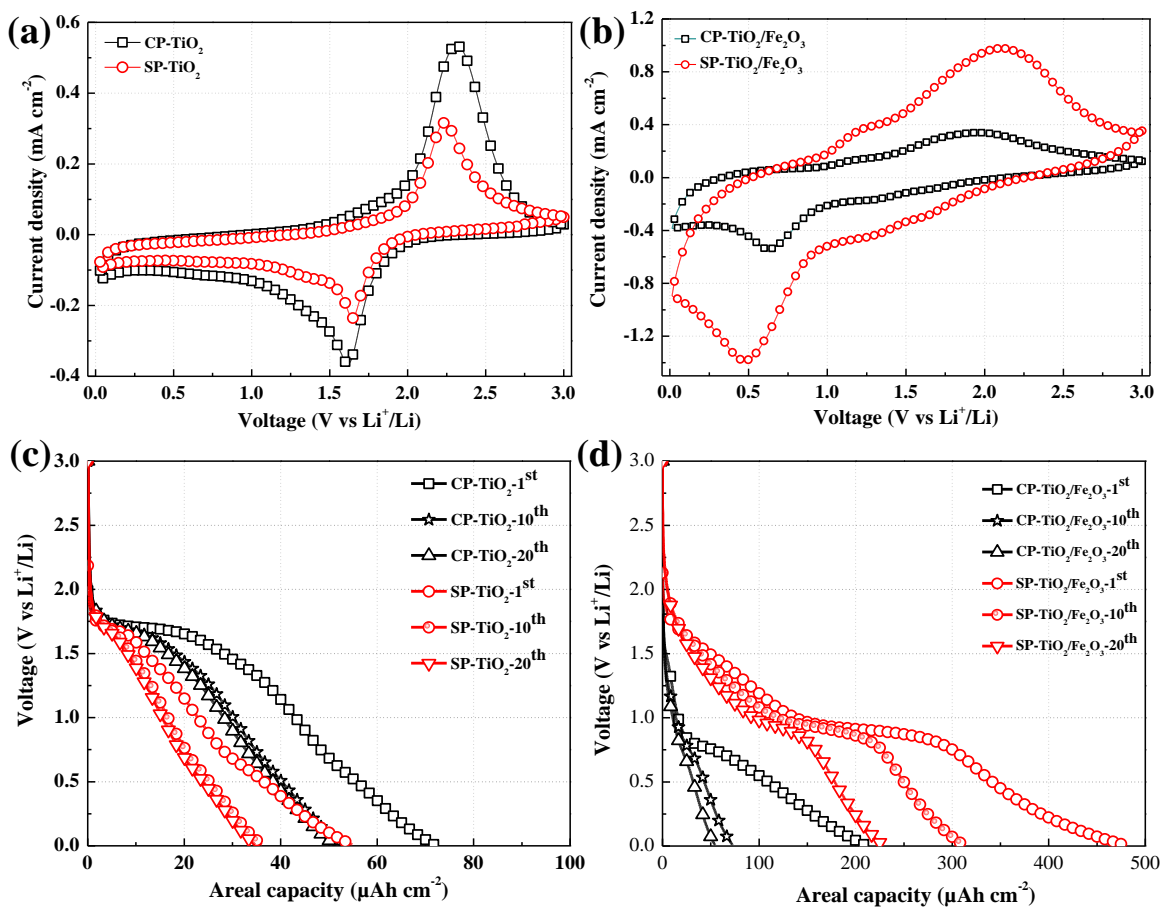

**Figure 5 Comparison of battery performance:** Cyclic voltammetry (CV) of (**a**) bare spaced (SP-TiO$_2$) and close-packed (CP-TiO$_2$) nanotubes. CV of (**b**) α-Fe$_2$O$_3$ decorated spaced (SP-TiO$_2$/Fe$_2$O$_3$) and close-packed (CP-TiO$_2$/Fe$_2$O$_3$) nanotubes. Comparison of galvanostatic discharge curves of (**c**) SP-TiO$_2$ and CP-TiO$_2$, (**d**) SP-TiO$_2$/Fe$_2$O$_3$ and CP-TiO$_2$/Fe$_2$O$_3$ nanotubes.

The separation between lithiation and delithiation peaks is lower for spaced NTs (i.e., 0.59 V) than close-packed NTs (i.e., 0.69 V), which indicates a faster Li$^+$ diffusion or lower overpotential, as reported previously [44]. In literature, peak to peak separation is reported as 0.45 V for anodic TiO$_2$ nanotubes [45], as 0.36 V for hydrothermal TiO$_2$ nanotubes [46], and as 0.28 V for 10 nm sized TiO$_2$ nanoparticles [44]. The large separation and the deviation from the Nernstian redox reactions (i.e., 59 mV for one electron reaction) are associated with slow electron transport, transport limitations in the electrolyte, a slow reaction kinetics [24,47,48], or can be due to semiconductor nature of TiO$_2$ [49].

We further evaluated the influence of Li$^+$ intercalation as well as deintercalation on the crystal structure of TiO$_2$ as shown in Figure S7. Once the intercalation process takes place, the anatase peak becomes smaller. Upon Li deintercalation a significant reduction in the peak intensity is observed. The reduction in the peak intensity (due to $Li^+$ intercalation as well as to deintercalation) can be due to the amorphization of $TiO_2$, as reported previously [34,50,51].

The influence of secondary metal oxide loading ($\alpha$-$Fe_2O_3$) on the LIB performance was investigated for both CP and SP NTs. Figure 5(b) and S8 show the CVs for the close-packed and spaced nanotubular layers decorated with different iron salt concentrations between 10 to 160 mM.

While for spaced NTs the current density increases significantly with increasing concentration (up to 120 mM) of Fe-precursor, in the case of close-packed NTs the current density barely increases and results in a comparable Li-battery performance as for bare close-packed $TiO_2$ NTs (see Figure S8). The findings for spaced nanotubes suggest that 120 mM provides an optimal loading with the current density reaching the peak point. This means that for the SP NTs, although they have less surface area than the CP tubes, the spacing enables an easy access for the deposition of a secondary metal oxide. Additionally, the spacing between NTs promotes electrolyte penetration during $Li^+$ insertion/extraction. The cathodic peak at about 0.47 V indicates a complete reduction of $Fe^{+3}$ to $Fe^0$ and the formation of $Li_2O$. The wide cathodic peak may be an indication of the formation of an irreversible solid-electrolyte interphase (SEI) formation. In the anodic polarization/sweep process (i.e., delithiation), a wide peak (composed of two peaks) was recorded at about 2.1 V corresponding to the oxidation of $Fe^0$ to $Fe^{+3}$ and delithiation of $TiO_2$, respectively (Figure 5(b)), as described in literature [52].

Figure 5(c)–(d) illustrate the galvanostatic discharge curves for bare and $\alpha$-$Fe_2O_3$ decorated nanotubes. In the galvanostatic discharge cycle of bare close-packed and spaced NTs, the potential rapidly drops to reach a well-defined plateau at 1.75 V associated with the intercalation of $Li^+$ through a two-phase equilibrium of a Li-poor (tetragonal) and a Li-rich (orthorhombic) phase, see Figure 5(c). In previous works, a Li intercalation plateau is observed at ~1.7 V and at ~1.8 V for anodic $TiO_2$ NTs [53,54], at ~1.76 V for hydrothermally grown $TiO_2$ NTs [46] and at 1.8 V for anatase $TiO_2$ nanoparticles [55]. In the case of iron decorated nanotubes, a broad band between 1.5 and 0.7 V with a maximum at ~1 V is visible, meaning that $Fe^{3+}$ ions are reduced to $Fe^0$ (Figure 5d), in line with literature [50].

The voltage (vs. $Li/Li^+$) versus areal capacity graph shows that the capacity calculated from the 1$^{st}$ discharge cycle for close-packed and spaced bare NTs are 71 µAh cm$^{-2}$ (~51 µAh cm$^{-2}$ for 20$^{th}$ discharge cycle), 54 µAh cm$^{-2}$ (~34 µAh cm$^{-2}$ for 20$^{th}$ discharge cycle) at the rate of 0.2 mA cm$^{-2}$, respectively (Figure 5(c)). The reported capacitance values for $TiO_2$ vary depending on the morphology, crystal structure and further treatments. Wei et al. [56] produced 4.5, 9 and 14.5 µm long smooth NTs using two-step anodization. These 9 µm smooth NTs provide high and stable areal capacities of 0.24 mAh/cm$^2$ at a charge/discharge current density of 2.5 mA/cm$^2$. For longer nanotubes e.g. 14.5 µm, an areal capacity of 0.70 mAh/cm$^2$ was obtained in the first cycles, which decreased to 0.60 mAh/cm$^2$ after 100 cycles.

The α-Fe$_2$O$_3$ loaded spaced TiO$_2$ nanotubes show an areal capacitance of 477 µAh cm$^{-2}$ – that is an approximately ~8 times increase compared with close-packed NTs (208 µAh cm$^{-2}$, limited to ~3 times increase, note that 1$^{st}$ cycles are considered for comparison), see Figure 5(d). The increase in capacitance is attributed to the fact that spaced NTs not only lead to good amount of metal oxide loading but also promote electrolyte penetration and better use of the active surface area for Li-ion intercalation.

Here, we observe a comparable performance for α-Fe$_2$O$_3$ loaded TiO$_2$ nanotubes with previously published results. For instance, Ortiz et al. [50,57] demonstrated titania nanotube and iron oxide nanowire composites negative electrode that is built by combining an anodic TiO$_2$ layer with electrodeposited Fe$_2$O$_3$ nanowires. The 3 µm thick nanocomposite electrode exhibited areal capacities of 468 mAh/cm$^2$ (specific capacities of 1190 mAh/g) at a rate of 25 mA/cm$^2$. Yu et al. [36] reported 600 µAh/cm$^2$ at a current density of 50 mA/cm$^2$ for 10 µm TiO$_2$ NTs decorated with Fe$_2$O$_3$ nanoparticles.

The bare spaced and close-packed NTs show good cyclic stability, i.e. close-packed NTs retain 84% and spaced NTs 53% of their initial capacities for a higher number of cycles. On the other hand, Fe$_2$O$_3$ decorated TiO$_2$ NTs show poorer cyclic stability due to Fe$_2$O$_3$ dominated electrochemical response of Fe$_2$O$_3$/TiO$_2$ electrodes. The poor cyclic performance of Fe$_2$O$_3$/TiO$_2$ electrodes is attributed to low electrical conductivity and iron oxide aggregation during charging/discharging, while TiO$_2$ NTs still maintain their morphology i.e., present no crack/fracture formation and show morphological stability.

## 4. Conclusion

The present study demonstrates the possibility to decorate the inner and outer walls of spaced TiO$_2$ nanotubular arrays (i.e. nanotubes with tube to tube spacing) with α-Fe$_2$O$_3$ nano-needles, and thus enhance their functional features. For this, the loading of spaced and close-packed nanotubes at different Fe-precursor concentrations are evaluated and optimum loading is achieved.

Overall, a spaced nanotubular TiO$_2$ array with optimum α-Fe$_2$O$_3$ (SP-TiO$_2$/Fe$_2$O$_3$) loading leads to improved Li-ion battery performance. I.e., the bare spaced NTs provide a 54 µAs cm$^{-2}$ capacitance, the functionalized electrodes (SP-TiO$_2$/Fe$_2$O$_3$), decorated with an optimized loading yield an areal capacitance of 477 µAh cm$^{-2}$. On the other hand, classic close-packed

NTs with α-Fe$_2$O$_3$ decoration (CP-TiO$_2$/Fe$_2$O$_3$) show an areal capacitance of 208 µAh cm$^{-2}$ – this value is restricted by the limited loading volume of this geometry. Furthermore, we believe that this concept, to load spaced tubular configuration with a secondary active or synergistic material not only is beneficial for an improved Li-ion battery performance of TiO$_2$ structures but also can find wider applications in further functional hierarchical structures.

**Acknowledgements**

The authors acknowledge the ERC, the DFG, the DFG "Engineering of Advanced Materials" cluster of excellence and DFG "funCOS" for financial support. Nhat Truong Nguyen is acknowledged for his help with the experiments.